\documentstyle[12pt]{article}
\begin{document}
\large
\vskip 2cm
\begin{center}
{ \Large {\bf On the Dirac equation for a quark
}}\\ \vspace*{1cm} {\bf Ivanhoe B.Pestov}\\ \vspace*{1cm}
{\it Joint Institute for Nuclear Research \\ Bogoliubov Laboratory of
 Theoretical Physics \\ 141980, Dubna, Russia} \\ {e-mail:
  pestov@thsun1.jinr.ru} \end{center}
\begin{abstract}
It is argued from geometrical, group-theoretical and physical points
of view that in the framework of QCD it is not only necessary but also
possible to modify the Dirac  equation so that correspondence principle
holds valid.  The Dirac wave equation for a quark is proposed and some
consequences are considered. In particular, it is shown that interquark
potential expresses the Coulomb law for the quarks and, in fact, coincides
with the known Cornell potential. \end{abstract}

\section{Introduction} According to the modern standpoint [1], spacetime
theory is any one that possesses a mathematical representation whose
elements are a smooth four-dimensional manifold $M$ and geometrical objects
defined on $M.$ The system of real local coordinates on $M$ is defined as a
topological mapping of an open region $U \subset M$ onto the Euclidean
four-dimensional space $ E_4.$ Thus, the Euclidean four-dimensional space
$E_4$ is a fundamental structure element of the mathematical formalism of
contemporary physics. However, it can be shown that $E_4$ has an underlying
structure that is exhibited in the existence of  group of transformation
that does not coincide with the $SO(4)$ group.  In physical space
$E_3$ - and, for the sake of comparison, on physical plane $E_2$- we
consider the groups of rotations and dilatations with the generators $$D =
x \frac{\partial}{\partial x} +y \frac{\partial}{\partial y} + z
\frac{\partial}{\partial z},\quad M_1= z \frac{\partial}{\partial y} - y
\frac{\partial}{\partial z} $$ $$ M_2 = z \frac{\partial}{\partial x} - x
\frac{\partial}{\partial z} ,\quad M_3 = -y \frac{\partial}{\partial x} + x
\frac{\partial}{\partial y}$$

in  $E_3$ and with generators $$D = x
\frac{\partial}{\partial x} +y
\frac{\partial}{\partial y} , \quad M = -y\frac{\partial}{\partial x}
+x \frac{\partial}{\partial y} $$ in $E_2.$ We denote these groups by $D
\otimes SO(3)$ and $D \otimes SO(2),$ respectively. It  can be shown that
  an element of first group can be parametrized by the real numbers
  $a,b,c,d ;$ it is suitable to consider them as a quaternion
$q= a i + b j  + c k + d .$ For the second group, we have two real parameters
combined into a complex number $w= u+iv.$ It is easy to verify one- to- one
correspondence between the algebras quaternions and complex numbers and the
 $D \otimes SO(3)$ and $D \otimes SO(2)$ groups. The transformations of the
 $D \otimes SO(3)$ and $D \otimes SO(2)$ groups in $E_3$ and $E_2$  can
be represented as  $$ R' = qR \bar q \quad r' = wrw,$$ where $
R=xi+yj+zk $ in the first case and $r =x+iy$ in the second case.  When we
consider  $D \otimes SO(3)$ and $D \otimes SO(2)$ groups as linear spaces
it can easily be seen, with the aid of the well-known algebra, that
the four- dimensional space of quaternions $Q_4$ and the two- dimensional
space of complex numbers, $Q_2,$ give spinor representations of the groups
in question. These representations are realized as follows:
\begin{equation} u' = qu, \quad z' = wz, \end{equation} where
$$u= u_1 i + u_2 j  + u_3 k + u_4, \quad  z = z_1 + z_2 i.$$
A remarkable
property of this transformations is that there is only one point $u=0, \,
(z=0)$ that is a stable  under the transformations given by (1).  When we
fix any other point, the transformations reduce to the
identity transformation.  Another important feature of this transformations
(1) is that the Euclidean scalar products in $Q_4$ and $Q_2$ $$(u,u)= u
\bar u = u_1^2 + u_2^2 + u_3^2 + u_4^2, \quad (z,z)= z \bar z = z_1^2 +
z_2^2 $$ are invariant under transformations (1), provided that $q\bar q =1
$ and $ w\bar w =1.$ But this does not mean that $Q_4$ and $Q_2$ are
Euclidean spaces because (1) hold. It should be noted, that there is a
simple mapping from $Q_2$ to $E_2$   of the form $$ r=z^2.$$ This mapping
is known as the Bolin transformation.  It can be shown, however,  that
there is no mapping from $Q_4$ to four-dimensional Euclidean space.  Thus,
what we usually call four-dimensional Euclidean by the analogy with a
three-dimensional physical space  is in fact $Q_4.$ That there is no such
mapping follows from the fact that real Dirac matrices $\gamma_i,$
possessing the properties $$\gamma_i \gamma_j + \gamma_j \gamma_i = 2
\delta_{ij}, \quad i,j =1,2,3,4 $$ do not exist. However, there is mapping
from $Q_4$ to $E_3$ which can be defined as  $$R= ui\bar u.$$ This mapping
is known as Hopf mapping.  Therefore, the components of a quaternion $u$
are not observable in a direct way, but only through some expressions
constructed from these components.  In seance, this situation is similar to
that  with a wave function in quantum mechanics.  It is obvious that a
three-dimensional sphere in $Q_4,$  $$u_1^2 + u_2^2 + u_3^2 + u_4^2
=\rho^2$$ inherits the properties of the enveloping space.  In view of the
unusual properties of four-dimensional space and of, respectively, a three-
dimensional sphere $S^3,$ we consider the investigations associated  with
the last object.

From the geometrical point of view a three-dimensional sphere is a space of
constant positive curvature. The Kepler-Coulomb problem in this space has a
long history and was first investigated by Schr\"odinger [2]. The
symmetry properties of the Schr\"odinger equation for the
Kepler-Coulomb problem in a space of constant positive curvature was
analyzed by Higgs [3] and by Leemon [4].  On the other hand $S^3$ is the
configuration space for a Top.  The quantum-mechanical problem for the free
motion of a Top was investigated shortly after the creation of quantum
mechanics in its moderns form (see for example [5]).  It is obvious that
connections between these two directions of investigation are very
important. Moreover in the 1930s, it was emphasized by Casimir [6] that,
from the physical point of view, the notion of a rigid bogy is as
fundamental as the notion of material point.  At last, we would like to
emphasize that QCD is conceptually a simple theory and that its structure
is  determined solely by  symmetry principles.  However, there is no
connection between such important phenomena as confinement and quark-lepton
symmetry, on one hand, and the first principles of QCD, on the other hand.
Despite prolonged and complicated experiments, free quarks have not yet
been observed, but it is commonly believed that quarks are true elementary
particles like electrons.  Experimentalists gradually arrived at the
conclusion that the matter is not in the details of the experiments but
rather in the fundamental properties of matter, which were sought under
various assumptions.  For instance, it is hypothesized that quark
confinement can be explained by topological methods that have recently
 found still a wider use in physics.  Nevertheless, the most natural and
reliable approach to the problem of confinement must be sought in the
possibility of modifying the original Dirac equation to take into account
the unusual properties of the quarks.

Summarizing all the facts considered above, we conjecture that the
configuration space of a quark coinces with the configuration space
for a Top.  Within this conjecture, we
will derive the Dirac wave equation for a quark
and consider some of its properties.

\section{Formulation of the Problem}

The choices of underlying spacetime manifold can
be reasoning as followers.  Consider a conventional quantum-mechanical
operator of the 4-momentum with components

\begin{equation}
P_{0}=-i\hbar \frac{\partial}{\partial x^{0}},\quad
P_{1}=-i\hbar \frac{\partial}{\partial x^{1}},\quad
P_{2}=-i\hbar \frac{\partial}{\partial x^{2}},\quad
P_{3}=-i\hbar \frac{\partial}{\partial x^{3}},
\end{equation}
where $x^{0}=ct,\quad x^{1}=x,\quad x^{2}=y,\quad x^{3}=z .$ Further, the
group of translations of spacetime is a finite continuous Lie group with
the generators

\begin{equation}
X_{0}=\frac{\partial}{\partial x^{0}},\quad
X_{1}=\frac{\partial}{\partial x^{1}},\quad
X_{2}=\frac{\partial}{\partial x^{2}},\quad
X_{3}=\frac{\partial}{\partial x^{3}}.
\end{equation}
Besides, the linear operators (3) are generators of a simply
transitive group of transformations [7]. In our case it is an Abelian
simply transitive group of transformations because
$$[X_{a},X_{b}]=0 ,\quad a,b=0,1,2,3. $$
The connection between(2) and (3) is obvious.
If a simply transitive group is a group of transformations of a
certain spacetime, this spacetime is called homogeneous, as this group can
transform any point of the spacetime to any a priori given point. To make
further analysis more transparent, we present the general
characteristic of  homogeneous spacetime manifolds.

As any vector field with components $V^{i}$ can be associated with
the linear operator $X=V^{i}\partial/{\partial x^{i}} $, the operator
$X$ is called the vector field. Then, let the vector fields  $X_a$
\begin{equation}
X_{a}=V^{i}_{a}\frac{\partial}{\partial x^{i}}  \quad (a=0,1,2,3)
\end{equation}
are generators of a simply transitive group of transformations of
a four-dimensional spacetime manifold $M$. Indexes from the
beginning of the Latin alfavet numerate vectors. In this case

\begin{equation}
[X_{a},X_{b}]= f^{c}_{ab} X_{c},
\end{equation}
where $f^{c}_{ab}$  are structure constants of the group in question. The
vector fields $V^{i}_{a}$ uniquely determine the system of covector fields
$V^{a}_{i}$ such that

\begin{equation}
V^{i}_{a} V^{a}_{j}=\delta^{i}_{j}, \quad
V^{a}_{i}V^{i}_{b}=\delta^{a}_{b}.
\end{equation}
The simply transitive group induces a natural integrable connection
$\Gamma$ on $M$ with Christoffel symbols

\begin{equation}
\Gamma ^{i}_{jk}=V^{i}_{a} \partial_{j} V^{a}_{k}
\end{equation}
and a natural metrics of the Lorentz signature on $M$

\begin{equation}
g_{ij}=\eta_{ab}V^{a}_{i}V^{b}_{j} ,\quad
g^{ij}=\eta^{ab}V^{i}_{a}V^{j}_{b},
\end{equation}
where $\eta_{ab}=\eta^{ab}=\rm diag(1,-1,-1,-1)$. From (5) and (7)
for the torsion tensor and torsion covector of the connection (7) we
obtain

$$T^{i}_{jk}=\Gamma^{i}_{jk}-\Gamma^{i}_{kj}=
-f^{a}_{bc}V^{i}_{a}V^{b}_{j}V^{c}_{k},\quad
T_{i}=T^{k}_{ki}=-f^{a}_{ab}V^{b}_{i}.$$
From (5) and (8) we have

\begin{equation}
 V^{i}_{a}V^{j}_{b;i}=
(f^{c}_{ab}-\eta_{ad}f^{d}_{bc}\eta^{ce}-
\eta_{bd}f^{d}_{ae}\eta^{ce})V^{j}_{c},
\end{equation}
where semicolon means the covariant derivative with respect to the
Levi-Civita connection of the metrics (8) with the Christoffel
symbols

$$\{^{i}_{jk}\}=\frac{1}{2}g^{il}(\partial_{j} g_{kl}+
\partial_{k} g_{jl}-\partial_{l} g_{jk}).$$
Knowing generators of the simply transitive group $X_{a}$ we can find
generators $Y_{a}$ of the mutual simply transitive group by solving
the equation $\nabla_{i} V^{j}- T^{j}_{ik} V^{k}=0 $, where
$\nabla_i$ is the covariant derivative with respect to the
connection (7).  For $X_{a},\quad Y_{a}$ we have $[X_{a},Y_{b}]=0
,\quad a,b=0,1,2,3 $.

The manifold $M$ that admits a simply transitive group of
transformations and has the metrics (8) will be called the
homogeneous spacetime manifold. The Minkowski spacetime is a particular
case of homogeneous spacetime manifolds. If a homogeneous spacetime
defined by a non-Abelian simply transitive group of transformations
has a physical meaning, from (2) and (3) it follows that the operator

$$-i\hbar X_{0}=-i\hbar V^{i}_{0}\frac{\partial}{\partial x^{i}}$$
will be analog of the operator $P_0$ in the Minkowski spacetime.

\section{The Dirac wave equation for a quark}

Here we will define a homogeneous
spacetime manifold that differs from the Minkowski spacetime by
geometrical and topological properties and show that a spacetime
manifold of that kind obeys all the required conditions and is of
definite interest for the physics of quarks.

In the five-dimensional Minkowski spacetime $M^5_{1,4} $ with
Cartesian coordinates $x^A$ (indices denoted by capital letters run
 through the five values $0,1,2,3,4$) and metrics

$$
ds^2 = \eta_{AB} dx^A dx^B =
(dx^0)^2 - (dx^1)^2 -(dx^2)^2 - (dx^3)^2 -(dx^4)^2 ,
$$
we will consider the one sheet hyperboloid  $ H^4 $

\begin{equation}
\eta_{AB}x^A x^B =
(x^0)^2 -(x^1)^2 -(x^2)^2 -(x^3)^2 -(x^4)^2 = - a^2,
\end{equation}
where $a$ is the radius of $H^4$, and prove that it is a homogeneous
spacetime manifold.

We will use the scalar product $(X,Y)=\eta_{AB}U^AV^B$ for any vector
fields $X=U^A\partial_A $ and  $Y=V^A\partial_A $  on $M^5_{1,4}$.
The vector fields
$$P_A =\delta^C_A \partial_C,\quad
M_{AB}=(x_A \delta^C_B -x_B\delta^C_A)\partial_C ,$$
where $x_A =\eta_{AB}x^B $, are generators of the Poincare group of
the five-dimensional Minkowski spacetime. All vectors fields
$M_{AB}$ are orthogonal to the radius-vector $R =x^C\partial_C $,  but this
is not the case for the vector fields $P_A.$   Representing $P_A$ as the
sum of the component aligned with the direction of the radius vector $R$
and the component orthogonal to this direction, we obtain the vector fields
$$M_A =aP_A +\frac{1}{a}(R,P_A)R= (a\delta^C_A
+\frac{1}{a}x_Ax^C)\partial_C, $$ which are tangent to $H^4$, because from
(10), it follows that $(R,M_A)=0$ at each point of $H^4$.  The vector
fields $M_A$ and $M_{AB}$ are generators of the group of conformal
transformations of $H^4$ because we have

\begin{equation}
[M_A,M_B] = -M_{AB}  ,\quad [M_A, M_{BC}] =
  \eta_{AB}M_C -\eta_{AC}M_B.
\end{equation}

Let us now introduce the vector fields

\begin{equation}
X_0 =M_0 ,\quad X_1 = M_{14} +M_{23} ,\quad
X_2 = M_{24}+ M_{31} ,\quad X_3 = M_{34}+M_{12}
\end{equation}
with the components

$$\begin{array}{l} X_0 =(a+\frac{x^2_0}{a} ,\quad \frac{x_0x^1}{a}
,\quad \frac{x_0x^2}{a} ,\quad \frac{x_0x^3}{a} ,\quad
\frac{x_0x^4}{a}),
\\ X_1 =(0 ,\quad -x_4 ,\quad -x_3 ,\quad x_2 ,\quad x_1 ),
\\ X_2  =(0 ,\quad  x_3 ,\quad - x_4 ,\quad -x_1 ,\quad x_2 ),
\\  X_3  = (0 ,\quad -x_2 ,\quad x_1 ,\quad -x_4 , \quad x_3 ).
 \end{array} $$
It is straightforward to see that the vector fields $X_0,\quad X_1
,\quad X_2 ,$ and $ X_3 $ are continuous and do not vanish at any point
of $H^4$. Because $(X_a,X_b) =0 $ for $a \neq b,\quad  a,b =0,1,2,3 $ and

$$( X_0,X_0) =\quad -(X_1,X_1) = \quad-(X_2,X_2) =\quad -(X_3,X_3) =
\quad a^2 + x^2_0, $$
the vector fields $X_0,\quad X_1,\quad X_2,$ and $ X_3 $ are linearly
independent at each point of $H^4$. From (11), it follows that

$$[X_0,X_i] =0 ,\quad [X_i,X_j ] =2e_{ijk}X_k ,\quad i,j,k =1,2,3 ,$$
where $e_{ijk}$ is the completely antisymmetric Levi-Civita symbol
specified by the equality $e_{123}=1.$ In this way, we have proven that the
one sheet hyperboloid  (10) admits a simply transitive group of
transformations whose generators are given by (12) and which has only
the following nonzero structure constants:

 \begin{equation}
f^1_{23} =f^2_{31}= f^3_{12} =2 .
\end{equation}
Therefore, we will supply   $H^4$ with a metrics of the type (8) and
 thus transform $H^4$ into the hyperbolic spacetime $H^4_{1,3}.$
From (9) and (13) it follows that the vector field $X_0$ is
absolutely parallel with respect to the Levi-Civita connection on
$H^4_{1,3}$ induced by the vector fields (12). For comparison we note
that the vector field $X_0=\partial / {\partial x^0} $ defined in (2)
is also absolutely parallel.
Now it is natural to put forward the idea that in the realm of the
strong interactions spacetime geometrically can be represented as
a one-sheeted hyperboloid (10)
in the five-dimensional  Minkowski spacetime.   A constant $a$ can be
interpreted geometrically as the radius of the three-dimensional sphere
$S^3,$ which is considered here as a space section $x^0 = 0.$ From the
physical point of view we treat  $a$ as the size of region of quark
confinement, since a quark is a pointlike particle in the space of constant
positive curvature $S^3.$ For comparison of leptons and quarks we note that
free motion of the electron is represented as a straight line in the
Euclidean usual space and free motion of the quark is a circumference on
the 3d sphere.  This correlation between leptons and quarks will be
continued with an example of the Coulomb law for these objects.    To do
this consider the Dirac equation for a quark.

In accordance with the original Dirac equation, we write
the Dirac equation in the homogeneous spacetime in the form

\begin{equation}
\gamma^c P_c \psi = \mu \psi,
\end{equation}
where

$$\gamma^a \gamma^b + \gamma^b \gamma^a = - 2 \eta^{ab},$$

$$P_c = X_c + \frac{iqa}{\hbar c} {A_c} -\frac{1}{2} {f_c},\quad f_c
= f^a_{ac} . $$
Here $q$ is the charge of a particle, and $A_c$ are the components
of the vector potential of the electromagnetic field in the basis
$X_a.$ In $H^4_{1,3},$  we also have
$$ \mu = {mca}/{\hbar}.$$
For the time being, we do not specify
the values of the structure constants of a simply transitive group of
transformations of the spacetime $H^4_{1,3}.$  In general, $[X_a, X_b]  =
f^c_{ab} X_c;$ therefore, we have

$$[P_a, P_b] = f^c_{ab}P_c + \frac{iqa}{\hbar c} {F_{ab}} ,$$
where

\begin{equation}
F_{ab} = X_a A_b - X_b A_a - f^c_{ab} A_c
\end{equation}
are the components of the strength tensor of the electromagnetic field in
the basis $X_a$. When the wave equation is established it is not
difficult to write the equations of electromagnetic field. The Jacobi
 identity $[P_a[P_b, P_c]] + [P_b[P_c,P_a]] + [P_c[P_a,P_b]] =0 $
 results in the first four Maxwell equations

\begin{equation}
X_a F_{bc} + X_b F_{ca} + X_c F_{ab} +
f^d_{ab} F_{cd} + f^d_{bc} F_{ad} + f^d_{ca} F_{bd} =0
\end{equation}
To establish the form of other four Maxwell equations, we set $\tilde
F^{ab} = \frac{1}{2} e^{abcd} F_{cd}$, where $e^{abcd}$ are
components of the antisymmetric Levi-Civita unit tensor in the basis
$X_a$. Then we can write equations (16) in the following equivalent
form

\begin{equation}
X_a \tilde F^{ab} + f_a \tilde F^{ab} +
\frac{1}{2}f^b_{ad} \tilde F^{ad} = 0
\end{equation}
By analogy, from (17)  it follows that the remaining Maxwell
equations are of the form

\begin{equation}
X_a F^{ab} + f_a F^{ab} +
\frac{1}{2} {f^b_{ad}} F^{ad}  = \frac{4{\pi}a}{c} {j^b},
\end{equation}
where $j^b$ are components of the current vector in the basis $X_a.$

Now we will write the Maxwell equations in the three-dimensional
vector form.  As usual, we put

$$j^a = (c\rho ,{\rm \vec j}),\quad
A_a =(\varphi,\quad - \rm \vec A),$$

$$ E_i = F_{0i}, \quad H_i = \frac{1}{2} e_{ijk}F^{jk} ,\quad
i,j,k =1,2,3 .$$

Then from (13) and (15)  we obtain

\begin{equation}
\rm \vec E = - \nabla_0 \vec A - \nabla \varphi ,\quad
\vec H = rot \vec A = \nabla \times \vec A -2\vec A.
\end{equation}
where

$$\nabla = (\nabla_1,\quad \nabla_2, \quad \nabla_3 ) ,
\quad \nabla_0 = X_0, \quad \nabla_i = X_i,\quad i = 1,2,3.$$

Considering that ${\rm div }\vec A = \sum^{3}_{i=1} \nabla_i A_i $,
we can recast the Maxwell equations (17) and (18) into the familiar
vector form

\begin{equation}
\rm -\nabla_0 \vec H = rot \vec E ,\quad
\rm div \vec  H =0,\quad
\rm rot \vec H = \nabla_0 \vec E + \frac{4{\pi}a}{c}  {\vec j},\quad
\rm div \vec E = 4{\pi}a {\rho}.
\end{equation}

Making use of the commutation relations $[\nabla_i,\nabla_j]
=2e_{ijk} \nabla_k,\quad i,j,k =1,2,3,$ we can easily verify the identities

$$\rm div \,rot=0,\quad \rm rot \,grad = 0.$$
In addition, we have

$$\rm div \,grad= \triangle,$$
where $\triangle$ is the Laplacian on a three-dimensional sphere.
Torsion ( that is, the non- Abelian character of a simply transitive group
of transformations of $H^4$ ) manifests itself not only in the definition
of the operator $\rm rot$, (21), but also in the identity

$$(\rm rot +1)^2 = -\triangle + 1 + \rm grad div.$$

Since the space section of the $H^4_{1,3}$ is a three dimensional sphere,
it is interesting to show that the Dirac equation (14) is associated with
the Schrodinger equation for a spherical Top.
To verify this, we will derive
eigenvalues $E$ of the Dirac Hamiltonian in question when there is no
electromagnetic field that is, for  $F_{ab} = 0$. Squaring equation (14)
and using (13), we obtain the following equation for $E$

\begin{equation}
E^2 \psi = m^2 c^4 \psi - \frac{c^2 \hbar ^2}{a^2}
   {(\triangle + P)} \psi,
\end{equation}
where

$$P = \Sigma_1 \nabla_1 + \Sigma_2  \nabla_2 + \Sigma_3 \nabla_3 $$

and $\Sigma_i = \frac{1}{2} e_{ijk} \gamma^j \gamma^k .$ The operator
$P$ has properties analogous to those of the operator $ \rm rot$. In
particular, we have

\begin{equation}
(P + 1) ^2  =  - \triangle + 1
\end{equation}
Since $\triangle + P = -P(P+1),$  then
$$E^2  = m^2 c^4  +p(p+1) \frac{c^2 \hbar ^2}{a^2},$$
where $p$ is an egenvalue of the operator $P.$

To determine eigenvalues of the operator $P$, we consider Hermitian
operators acting in the space of solutions to the Dirac equation
(14). Generators of the group mutual to the simply transitive group
of transformations of the spacetime $H^4_{1,3}$ are given by

$$Y_0 = X_0 ,\quad Y_1 = M_{14} - M_{23},\quad
Y_2 = M_{24} - M_{31},\quad Y_3 = M_{34} - M_{12}.$$
This leads to the three Hermitian operators

$$N_i = - \frac{i}{2} Y_i $$
which are analogous to the momentum operators. The remaining three
operators,

\begin{equation}
M_i = - \frac {i}{2} (\nabla_i - \Sigma_i ) =
- \frac {i}{2}(X_i - \Sigma_i)
\end{equation}
are analogs of the electron angular momentum operators. From
(23), it follows that the spin of a particle in question is $\hbar / {2
}.$ We have

$$\rm \vec M \times \vec M = i \vec M ,\quad
\vec N \times \vec N = i \vec N $$
and, in addition,

\begin{equation}
2(M^2 + N^2) =  (P + \frac {3}{2})^2 -\frac{3}{4},
\quad 2(M^2 - N^2) = P + \frac{3}{2}.
\end{equation}
Hence, we have the operator equation
$$2(M^2 + N^2)+\frac{3}{4} =  4(M^2 - N^2)^2 .$$
Since, $M^2 = l(l+1), \quad N^2 = k(k+1),$ we find from the operator
equation that  $l$ and $k$  satisfy the equation
	  $$2[l(l+1) + k(k+1)]+\frac{3}{4} = 4[l(l+1) - k(k+1)]^2.$$
This equation has two solutions, $l=k+\frac{1}{2}$ and
$k=l+\frac{1}{2},$ whence it followers that
$p=2[l(l+1)-k(k+1)]- \frac{3}{2} = -2k-3.$  Since $S^3$ has a metric
invariant with respect to the isometrical reflection,  $p=2k+3$
is an eqenvalue too. For the energy, we then have
\begin{equation}
E^2 = m^2 c^4 + n(n+1) \frac {c^2 \hbar^{2}}{a^2} =
m^2c^4(1+\frac{\lambda^2}{a^2}) , \end{equation} where $n= 2,3,...$
and $\lambda = \hbar/mc.$   If formula (25) gives the
quantum-mechanical value of the energy of the relativistic spherical Top,
we can then conclude that, at large $a,$ the moment of inertia $I=ma^2$ is
also large, so that the angular velocity is small. Therefore, the
nonrelativistic limit can be found from the condition $a \gg \lambda .$ In
the limit of large $a$ it follows from (25) that

$$E = mc^2 + \frac{L^2}{2I},$$
where $L^2 =n(n+1)\hbar^2$ is the angular momentum of the
spherical top and $I$ is its moment of inertia.  The last formula is in
accordance with the classical formula

$$E = \frac {L^2}{2I} $$
for the energy of a Top.
Thus, the formula (25) gives energy of rotation.

Let us now consider the Coulomb law. The
Coulomb potential can be derived as a solution of the
equations of electrostatics, which are invariant under the group of
Euclidean motions, including rotations and translations. In the case being
considered, we will seek a Coulomb potential in
an analogous manner.  From (19) and (20), it follows that for a constant
electric field $\rm div \vec E = 4{\pi }a {\rho},\quad \vec E = - \nabla
\varphi,$ and consequently, $\varphi$ obeys the equation

\begin{equation}
\triangle \varphi = - 4 {\pi}{a^2} {\rho}.
\end{equation}

As it is well known, the electron Coulomb potential
$$\phi_{e}(r) = \frac{e}{r}$$ is the fundamental solution to the
Laplace equation $\triangle {\phi} = \mbox{div\, grad} \,{\phi} = o.
$ In accordance with our conjecture the Coulomb potential for quarks
can be derived as follows. Consider the stereographic projection
$S^3$ from point (0,0,0,-a) onto the ball $x^2+y^2+z^2 \leq a^2:$
$$x^1 = fx, \quad x^2 = f y, \quad x^3 = f z, \quad x^4 =
a(1-f),$$ where $f= 2a^2 /(a^2 + r^2).$  Then, it follows that the
element of length on the three-dimensional sphere can be represented in
the form $$ ds^2 = f^2 (dx^2 + dy^2 + dz^2)$$ and hence the Laplace
equation on $S^3$ can be written as follows $$\triangle \phi= f^{-3} \mbox
{div} (f \mbox{grad }\, \phi) = 0.$$ We seek the solution to this
 equation that is invariant under the transformation of the
 group $SO(3)$ with generators $$x^1 \frac{\partial}{\partial x^2} -
 x^2 \frac{\partial} {\partial x^1} =  x \frac{\partial}{\partial y}
 - y \frac{\partial}{\partial x} ,\quad etc.   $$   This subgroup of the
   $SO(4)$  group is determined by fixing the point $(0,0,0,-a).$ Let us
   put $$\psi = f \frac{1}{r} \frac{d\phi}{dr}.$$  Since $$\triangle \phi =
f^{-3}(r \frac{d\psi}{dr} + 3\psi) =  r^{-2} f^{-3}\frac{d}{dr}(r^3
   \psi),$$  then $r^3\psi = c_1 = constant.$ Thus, we have$$
\frac{d\phi}{dr} = c_1 \frac{a^2 + r^2}{2a^2r^2} = c_1
   (\frac{1}{2r^2} + \frac{1}{2a^2})$$  and hence
   \begin{equation}
		 \phi_q = c_1 ( -\frac{1}{2r} + \frac{r}{2a^2}) + c_2.
   \end{equation}
Generally speaking, expression (27), that we have derived, coincides with
the well-known Cornell potential [8],[9]. If we demand that $\phi_e(a)
= \phi_q(a),$ then $c_2= e/a $ and the Coulomb law for quarks has
the form \begin{equation} \phi_q(r) = q(\frac{1}{2r} - \frac{r}{2a^2})
			   + \frac{e}{a}, \end{equation} where $q$ is the
quark charge.

From our consideration it follows that the idea of
rotating matter can be realized in the framework of
relativistic quantum mechanics by the Dirac equation (14).

\section{Conclusion}

 As the basic wave equation describing the dynamics of quarks, we
have proposed the modified Dirac equation (14), which has been written here
in homogeneous coordinates. The conclusion that quarks are described by the
wave equation different from the conventional wave equation for electrons
is quite natural.  In fact, it would be strange if the description of such
different particles were based on the same equation.

 The physical meaning of the confinement phenomenon is tightly connected
with idea of a rotating matter and consists in that quarks possess
properties of a quantum-mechanical spherical Top.  Among other things, this
means that the interquark potential expresses the Coulomb law for quarks
and, in fact, coincides with the well-known Cornell potential that was
first very successfully used by the Cornell group.

For $a \rightarrow   \infty  ,$  the theory of electrons can be derived
from the theory of quarks, but the electrons are obviously deconfined
because in this case, the region of confinement covers the entire Euclidean
space. Thus, the symmetry between quarks and leptons has a natural
explanation.

Since the kinematics of quarks differs from the kinematics of
electrons, decays of hadrons and nuclei such that the energy
is conserved,  but the momentum is not, are possible.
Much has to be done, but the problem is worth efforts as many interesting
applications become possible.

\end{document}